\documentstyle[12pt,epsf]{article}
\begin{document}
\baselineskip=0.7cm
\newcommand{\EQ}{\begin{equation}}
\newcommand{\EN}{\end{equation}}
\newcommand{\EQA}{\begin{eqnarray}}
\newcommand{\EQN}{\end{eqnarray}}
\newcommand{\EQAN}{\begin{eqnarray*}}
\newcommand{\EQNN}{\end{eqnarray*}}
\newcommand{\e}{{\rm e}}
\renewcommand{\theequation}{\arabic{section}.\arabic{equation}}
\newcommand{\Tr}{{\rm Tr}}
\renewcommand{\thesection}{\arabic{section}.}
\renewcommand{\thesubsection}{\arabic{section}.\arabic{subsection}}
\makeatletter
\def\section{\@startsection{section}{1}{\z@}{-3.5ex plus -1ex minus
 -.2ex}{2.3ex plus .2ex}{\large}}
\def\subsection{\@startsection{subsection}{2}{\z@}
{-3.25ex plus -1ex minus
 -.2ex}{1.5ex plus .2ex}{\normalsize\it}}
\def\appendix{
\par
\setcounter{section}{0}
\setcounter{subsection}{0}
\def\thesection{\Alph{section}}}
\makeatother
\def\thefootnote{\fnsymbol{footnote}}
\begin{flushright}
hep-th/0010173\\
UT-KOMABA/2000-12\\
October, 2000
\end{flushright}
\vspace{1cm}
\begin{center}
\Large
Nonlinear Supersymmetry 
Without the 
GSO Projection \\
and Unstable D9-Brane
 
\vspace{1cm}
\normalsize
{\sc Tabito Hara}
\footnote{
e-mail address: {\tt tabito@hep1.c.u-tokyo.ac.jp}}
and 
{\sc Tamiaki Yoneya}
\footnote{
e-mail address: {\tt tam@hep1.c.u-tokyo.ac.jp}}

\vspace{0.3cm}

{\it Institute of Physics, University of Tokyo\\
Komaba, Meguro-ku, Tokyo 153-8902, Japan}

\vspace{1cm}
Abstract
\end{center}

Orientable open string theories containing both bosons 
and fermions  without the GSO projection are
expected to have the 10 dimensional $N=2$(A) space-time
supersymmetry in a spontaneously broken phase.   
We study the low-energy theorem 
for the nonlinearly realized $N=2$  supersymmetry using 
the effective action for an unstable D9-brane. 
It is explicitly confirmed that the 4-fermion open string
amplitudes without the GSO projection obey 
the low-energy theorem derived from the nonlinear $N=2$ 
supersymmetry. An intimate connection
between the existence of the hidden supersymmetry and the
open-open  string ($s$-$t$) duality  is pointed out.

\newpage
\section{Introduction}
In uncovering the whole structure of 
string theory, it seems important to understand not only 
perturbatively stable vacua but also some important classes 
of  unstable vacua.  In particular, the unstable vacua
corresponding to nonBPS D9($\overline{{\rm D9}}$)-branes
are considered to be of fundamental importance
\cite{sen0}\cite{witten}\cite{hora}. In connection with
this, the condensation of tachyon by which both types of
vacua can  be related to each other, has been a focus of
much interest recently. 
   From the viewpoint of symmetry structure, we expect that
one and the same supersymmetry must govern both the 
stable and unstable vacua in different ways. This is so even 
in the presence of tachyons signaling instability 
in the case of the unstable vacua. 
It is not, however, evident how the space-time supersymmetry 
is realized in string pertubation theory with tachyons,
and the existence of spontaneously broken 
supersymmetry in the presence of tachyons has never 
been explicitly proven.  We emphasize that this problem
is among several unanswered questions  concerning the
principles of string theory.  We should have some direct
formulation of supersymmetries within the intrinsic logic
of string theory. 

There are only 
a few works which studied the hidden supersymmetry in unstable
vacua.  On one hand, Sen \cite{sen} has proposed the effective
world-volume action for nonBPS D-branes in type II 
theories. The spontaneous
breaking of supersymmetry in this proposal is reflected to the
absence of $\kappa$-symmetry for the world-volume action. 
This leads to the correct degrees-of-freedom  of
massless   modes, in which the numbers of bosonic and 
fermionic  on-shell degrees of freedom are different, as
expected in general from  spontaneously broken
supersymmetry.   On the other hand, from the viewpoint of
open-string  theories which are supposed to provide the
exact (finite $\alpha'$) theory for stable and unstable
D-branes, it has been investigated
\cite{yo} by one of the present authors
 how the $N=2$ supersymmetry is buried in the
properties  of spectra, vertex operators, the  boundary 
conditions, etc, in both the NSR and GS formalisms.
Moreover, a formal construction  of the supersymmetry
transformation law is given, assuming the framework of
Witten's open string field  theory. 

The purpose of this note is to present further evidence 
for the hidden nonlinear supersymmetry without the 
ordinary GSO projection 
 as a support for the ideas 
discussed in the above two works. 
 We study the low-energy theorem for the scattering of would-be 
Goldstone fermions.
To the present authors' knowledge,  the fermion scattering of
open strings involving  both the GSO and opposite-GSO
projected sectors has not been  fully studied in the
literature.  We perform a detailed study of  the 
$s$-channel and
$t$-channel duality  properties of relevant fermion
amplitudes.  It  is shown that the $s$-$t$ duality  
is responsible for the emergence of the 
hidden nonlinear supersymmetry. 
The low-energy effective action for fermion scattering 
is constructed on the basis of Sen's proposal and compared 
to the zero-slope limit of the 4-point amplitudes of the 
Goldstone fermions. 

  In the next section, 
we construct the low-energy effective action for the 
10-dimensional open superstring theory with the 
$N=2$ space-time supersymmetry being realized 
nonlinearly, on the basis of the effective world-volume action 
for a space-time filling D9-brane. 
In section 3,  we study the structure of 
four-fermion scattering amplitudes involving 
both the GSO sectors, using the old results given by Schwarz
and Wu \cite{schwu} and paying special 
attention on their ($s$-$t$) duality properties.  The
zero-slope limit of the amplitudes is  studied. The result
agrees precisely with the behaviors predicted from the
supersymmetric effective action  and also from the arguments
of \cite{yo}.  The final section is devoted to concluding
remarks. 

\section{Low-energy effective action}

The work \cite{yo} gave strong evidence for believing that the
$N=2$  space-time supersymmetry is hidden in orientable 
open string theory containing both bosons and 
fermions without the GSO
projection.  In particular, all the
massive excitations are shown to participate in essential ways.
This means that if only the massless modes are  used, the
effective theories should exhibit the supersymmetry to 
all orders in
the derivative (or $\alpha'$-) expansion  in some nonlinear
realization. Unfortunately, however,  it is not easy to
derive the nonlinear realization directly  in this case,
since we do not know any complete off-shell superspace 
representation of the $N=2$ supersymmetry in 10
dimensions, containing the 
usual $N=1$ Maxwell (or Yang-Mills) supermultiplet. If
we restrict ourselves to  the low-energy limit,
however, we can switch to the  D9-brane picture in
type IIA theory, following the
procedure used originally in \cite{aps} for the BPS
 D9-brane of type IIB theory. For sufficiently
low energies,  we can assume the static gauge condition for
fixing the world-volume coordinates and derive the 
low-energy effective action for open strings in 10
dimensions  from the effective  world-volume action for
D9-brane which is believed to  describe the
low-energy behavior of open strings coupled to 
D9-brane.  The effective
action obtained in this way takes a form of the 
Volkov-Akulov type \cite{volaku}. 

Let us briefly recapitulate the
world-volume action  for unstable D-branes 
proposed in \cite{sen}. The action is nothing but
the Dirac-Born-Infeld part of the 
$\kappa$-symmetric action constructed in \cite{aps} for 
BPS Dp-branes ($\tau_p= 1/(g_s\sqrt{\alpha'}(2\pi
\sqrt{\alpha'})^p)$):
\EQ
S_p=-\tau_p \int d^{p+1}\sigma 
\sqrt{-\det(\cal{G}_{\mu\nu}+ 
{\cal F}_{\mu\nu})}  ,
\label{dpaction} 
\EN 
where 
\EQ
{\cal G}_{\mu\nu}= \eta_{mn}\Pi_{\mu}^m\Pi_{\nu}^n ,
\EN
\EQ
{\cal F}_{\mu\nu}=F_{\mu\nu} -
i[\overline{\theta}\Gamma_{11}\Gamma_m\partial_{\mu}\theta
(\partial_{\nu}X^m -{i\over
2}\overline{\theta}\Gamma^m\partial_{\nu}\theta)-
(\mu\leftrightarrow \nu)] ,
\EN
with 
\EQ
\Pi_{\mu}^m=\partial_{\mu}X^m -i\overline{\theta}\Gamma^m
\partial_{\mu}\theta  .
\EN
Our conventions for the metric and $\Gamma$
matrices are $\eta_{\mu\nu}=(+,+,\ldots, +, -), 
 \{\Gamma_{\mu}, 
\Gamma_{\nu}\}=2\eta_{\mu\nu}. $ 
The world volume indices are Greek $(\mu, \nu,
\ldots)$ and the space-time indices are 
lower-case alphabets $(m, n, \ldots)$. 
 Note that the spinor field
$\theta$  is a 10-dimensional Majorana spinor with 32
components.\footnote{
Note that our convention for the spinor field is 
slightly different from those of refs. \cite{aps} 
and \cite{sen} in that we put the factor $i=
\sqrt{-1}$ in front of $\overline{\theta}\Gamma_{\mu}
\partial_{\nu}\theta$ to fit to the standard 
field theory convention.
}  We denote the Weyl components by
$\theta_{\pm}$  which satisfy $\Gamma_{11}\theta_{\pm}=\mp
\theta_{\pm}, 
\overline{\theta}_{\pm}\Gamma_{11}=\pm \overline{\theta}_{\pm}$. 
This action has an $N=2$ supersymmetry since the tensors 
${\cal G}_{\mu\nu}, {\cal F}_{\mu\nu}$ are separately invariant 
under the supertransformation
\EQ
\delta \theta = \epsilon, \quad \delta X^m
=i\overline{\epsilon}
\Gamma^m\theta,
\label{orgstr1}
\EN
\EQ
\delta A_{\mu}=i\overline{\epsilon}\Gamma_{11}\Gamma_m
\theta \partial_{\mu}X^m +
{1\over 6}(\overline{\epsilon}\Gamma_{11}\Gamma_m\theta
\overline{\theta}\Gamma^m\partial_{\mu}\theta 
+ \overline{\epsilon}\Gamma_m\theta
\overline{\theta}\Gamma_{11}\Gamma^m\partial_{\mu}\theta ) .
\label{orgstr2}
\EN

We now consider a space-time filling D9-brane in type IIA theory.
Then, in the low-energy limit, it is legitimate to set 
the world-volume and space-time coordinates equal, 
$\sigma^{\mu}\rightarrow x^{\mu}=X^m$ ($\mu=m,
\ldots)$,  by assuming that configurations with various 
types of folding do not contribute in the low-energy limit. 
\EQ
{\cal G}_{\mu\nu}=\eta_{\mu\nu} -
i\overline{\theta}
(\Gamma_{\mu}\partial_{\nu} +
\Gamma_{\nu}\partial_{\mu})\theta 
- \overline{\theta}\Gamma^{\alpha}\partial_{\mu}
\theta \overline{\theta}\Gamma_{\alpha}\partial_{\nu}\theta
,
\EN
\EQ
{\cal F}_{\mu\nu}=F_{\mu\nu}
-i[\overline{\theta}\Gamma_{11}\Gamma_{\alpha}\partial_{\mu}\theta
(\delta^{\alpha}_{\nu} -{i\over
2}\overline{\theta}\Gamma^{\alpha}\partial_{\nu}\theta)-
(\mu\leftrightarrow \nu)] .
\EN
Then we have 
\EQ
G_{\mu\nu} \equiv {\cal G}_{\mu\nu}+{\cal F}_{\mu\nu}
= \eta_{\mu\nu} + \lambda F_{\mu\nu} +
\lambda^2 S_{\mu\nu}^{(2)} + \lambda^4 S_{\mu\nu}^{(4)} ,
\EN
\EQ
S_{\mu\nu}^{(2)}= -i\overline{\psi}_+\Gamma_{\nu}\partial_{\mu}
\psi_+ - 
i\overline{\psi}_-\Gamma_{\mu}\partial_{\nu}
\psi_- ,
\EN
\EQ
S_{\mu\nu}^{(4)}=-{1\over 4}(\overline{\psi}_+\Gamma^{\alpha}
\partial_{\mu}\psi_+ \overline{\psi}\Gamma_{\alpha}\partial_{\nu}
\psi 
+
\overline{\psi}_-\Gamma^{\alpha}
\partial_{\nu}\psi_- \overline{\psi}\Gamma_{\alpha}\partial_{\mu}
\psi) .
\EN
Here we changed the normalization of the 
fields by resetting as $A_{\mu}
\rightarrow \lambda A_{\mu}, \theta = \lambda \psi/\sqrt{2}
=\lambda(\psi_+ +\psi_-)/\sqrt{2}$, \footnote{
Throughout the present paper, we always use 
32$\times$32 components notation for $\Gamma$ 
matrices, so that $\psi_{\pm}\equiv (1\mp \Gamma_{11})
\psi/2$. }
such that the final form of the effective action has the
standard normalization  of field theory. Note also that the 
fermion bilinears without chirality indices $\pm$ are the 
sum of two chiral sectors. 
The constant 
$\lambda$ is the Yang-Mills  coupling constant in 10
dimensions and is equal to the   square root of the inverse
tension: $\lambda = 1/\sqrt{\tau}_9
=(g_s(2\pi)^9\alpha'^5)^{1/2}$.  

The supertransformation law is now given as ($\epsilon 
\rightarrow \epsilon/\sqrt{2}$) 
\EQ
\delta \psi = {1\over \lambda}\epsilon
-i{\lambda\over 2}(\overline{\epsilon}\Gamma^{\mu}\psi)
\partial_{\mu}\psi, 
\label{strpsi}
\EN
\EQ
\delta A_{\mu}={i\over
2}\overline{\epsilon}\Gamma_{11}\Gamma_{\mu}
\psi  +
{\lambda^2\over
24}(\overline{\epsilon}\Gamma_{11}\Gamma_{\nu}\psi
\overline{\psi}\Gamma^{\nu}\partial_{\mu}\psi
+ \overline{\epsilon}\Gamma_{\nu}\psi
\overline{\psi}\Gamma_{11}\Gamma^{\nu}\partial_{\mu}\psi )
-i{\lambda \over 2}(\overline{\epsilon}\Gamma^{\nu}\psi)
\partial_{\nu}A_{\mu}
-i{\lambda \over 2}(\overline{\epsilon}\Gamma^{\nu}
\partial_{\mu}\psi)
A_{\nu} .
\label{stra}
\EN
Note that, in both equations (\ref{strpsi}) and (\ref{stra}), 
the last (two, in case of (\ref{stra}),) 
terms are originated from the 
reparametrization of the world volume,  
which is required to compensate the supertransformation of 
the D-brane coordinates as given in (\ref{orgstr1}) 
in preserving the static gauge condition
$\sigma^{\mu}=x^{\mu}=X^{\mu}$,  and also that in
(\ref{stra}) the right hand side is  gauge invariant up to
the field dependent gauge  transformation
$\partial_{\mu}\Lambda$  with $\Lambda=-i{\lambda\over 2}
(\overline{\epsilon}\Gamma^{\nu}\psi)
A_{\nu}$. 
The effective action 
\EQ
S_{{\it eff}}=-{1\over \lambda^2}\int d^{10}x 
\sqrt{-\det G_{\mu\nu}}
\label{effaction}
\EN
and the supertransformation law take quite similar forms as 
those of the 
well known Volkov-Akulov action \cite{volaku} 
for the simplest nonlinear realization of 
supersymmetry, except for the 
presence of the gauge field and the associated quartic 
fermion terms in $G_{\mu\nu}$.  
As a matter-of-fact,  
we should in general expect higher derivative 
corrections to this simplest form of the action, being the 
effective theory for interacting open strings.
However, at least in the  lowest nontrivial approximation with
respect to  the derivative expansion, we expect that 
this action describes the scattering
amplitudes of massless Ramond fermions if the open string
theory indeed has the hidden supersymmetry without the 
GSO projection and if the massless Ramond fermions behave
as the Goldstone fermions. In this sense, we can regard the
above action as the low-energy theorem for the
spontaneously broken
$N=2$ supersymmetry which should be buried in open  string
perturbation theory. 

By expanding the action (\ref{effaction}) up to the 
order $\lambda^2$, we obtain
\[
S_{{\it eff}}+{1\over \lambda^2}\int d^{10}x= 
\int d^{10}x \, 
\Big(-{1\over 4}F_{\mu\nu}^2 +{i\over 2}(\overline{\psi}_+
\Gamma^{\mu}\partial_{\mu}\psi_+ +
\overline{\psi}_-
\Gamma^{\mu}\partial_{\mu}\psi_-) \]
\[
-
{\lambda^2\over 32}(F_{\mu\nu}^2)^2
+{\lambda^2\over 8}F^{\mu\nu}F_{\nu\alpha}F^{\alpha\beta}
F_{\beta\mu} 
-i{\lambda\over 2} (\overline{\psi}_+\Gamma_{\nu}\partial_{\mu}
\psi_+ + 
\overline{\psi}_-\Gamma_{\mu}\partial_{\nu}
\psi_-)F^{\nu\mu}\]
\[
+i{\lambda^2\over 2}(\overline{\psi}_+
\Gamma_{\mu}\partial_{\nu}\psi_+ +
\overline{\psi}_-
\Gamma_{\nu}\partial_{\mu}\psi_-)F^{\mu\alpha}F_{\alpha}^{
\,\, \nu}
+i{\lambda^2\over 8}\overline{\psi}\Gamma^{\mu}\partial_{\mu}
\psi F^2\]
\[
+{\lambda^2\over 8}(\overline{\psi}_+
\Gamma^{\mu}\partial_{\mu}\psi_+ +
\overline{\psi}_-
\Gamma^{\mu}\partial_{\mu}\psi_-)^2
\]
\[
+{\lambda^2\over 8}(\overline{\psi}_+\Gamma_{\mu}\partial_{\nu}
\psi_+\overline{\psi}_+\Gamma^{\mu}\partial^{\nu}\psi_+
+\overline{\psi}_-\Gamma_{\mu}\partial_{\nu}
\psi_-\overline{\psi}_-\Gamma^{\mu}\partial^{\nu}\psi_-
)\]
\[
-{\lambda^2\over 4}(\overline{\psi}_+\Gamma_{\mu}\partial_{\nu}
\psi_+\overline{\psi}_+\Gamma^{\nu}\partial^{\mu}\psi_+
+\overline{\psi}_-\Gamma_{\mu}\partial_{\nu}
\psi_-\overline{\psi}_-\Gamma^{\nu}\partial^{\mu}\psi_-
)\]
\EQ
-{\lambda^2\over 4}\overline{\psi}_+\Gamma_{\mu}\partial_{\nu}
\psi_+\overline{\psi}_-\Gamma^{\mu}\partial^{\nu}\psi_-
+ \mbox{higher orders}
\Big)
\EN

This action reduces to the standard free action 
in the limit $\lambda \rightarrow 0$, while the 
supertransformation law given by (\ref{strpsi}) and 
(\ref{stra}) does not apparently reduce to the free 
form. In fact, if one rescales the supertransformation 
parameter by $\epsilon \rightarrow \lambda \epsilon$, 
the $\lambda\rightarrow 0$ limit of the 
supertransformation is simply $\delta \psi =\epsilon, 
\delta A_{\mu}=0$. But then the superalgebra 
becomes trivial. To preserve the superalgebra, 
it is not allowed to take the $\lambda\rightarrow 0$ 
limit in this way, since the cancellation in the 
product of the order
$1/\lambda$ and $\lambda$ terms give the 
correct superalgebra in terms of the original unscaled 
superparameter. 

The form of the action and, correspondingly, the 
supertransformation law are not unique, since 
there is the ambiguity of field redefinition. 
The field redefinition for 
spinor fields in 10 dimensions was studied in ref.
\cite{metrah}.  For the purposes of self-containedness 
of our exposition and of
making some small corrections to the latter reference, we
present general formulae.  Let us define a general form of
the action containing  quartic fermion terms. Here  we
consider each   Majorana-Weyl spinor, namely, either
$\psi_+$ or $\psi_-$,  separately, 
although we use the notation $\psi$ without subscript 
$\pm$. 
\[{\cal L}=-{1\over 4}F_{\mu\nu}^2 +{i\over 2}\overline{\psi}
\Gamma^{\mu}\partial_{\mu}\psi 
+a_1 \lambda^2 (F^2)^2 +a_2\lambda^2 (F)^4
\]
\[
+b_0 i\lambda \overline{\psi}\Gamma_{\mu}\partial_{\nu}\psi
F^{\mu\nu} 
+b_1i\lambda^2 \overline{\psi}\Gamma_{\mu\nu\rho}\psi
\partial_{\sigma} F^{\sigma\nu}F^{\mu\rho}+
b_2i\lambda^2\overline{\psi}\Gamma_{\mu\nu\rho}\psi
F^{\mu\sigma}\partial_{\sigma}F^{\nu\rho}
\]
\[
+b_3i\lambda^2 \overline{\psi}\Gamma^{\mu\nu\rho\sigma\tau}
\partial_{\tau}\psi F_{\mu\nu}F_{\rho\sigma}
+b_4i\lambda^2\overline{\psi}\Gamma_{\mu}\partial_{\nu}\psi
 F^{\mu\sigma}F_{\sigma}^{\, \, \, \nu}
+b_5i\lambda^2 \overline{\psi}\Gamma^{\mu}\partial_{\mu}\psi
F^2
\]
\[
+c_1\lambda^2\,
(\overline{\psi}\Gamma^{\mu}\partial_{\mu}\psi)^2
+c_2 \lambda^2\, \overline{\psi}\Gamma^{\mu}\partial^{\nu}\psi
\overline{\psi}\Gamma_{\mu}\partial_{\nu}\psi
+c_3\lambda^2 \overline{\psi}\Gamma^{\mu}\partial^{\nu}\psi
\overline{\psi}\Gamma_{\nu}\partial_{\mu}\psi\]
\EQ
+c_4 \lambda^2
\overline{\psi}\Gamma^{\mu\nu\rho}\partial_{\rho}\psi
\overline{\psi}\Gamma_{\mu}\partial_{\nu}\psi
+c_5\lambda^2
\overline{\psi}\Gamma^{\mu\nu\rho}\partial_{\rho}\psi
\overline{\psi}\Gamma_{\mu\nu\tau}\partial^{\tau}\psi .
\label{expandedaction1}
\EN
The most general form, to this order of $\lambda$-expansion, of
the field redefinition which preserves  chirality is 
\[
\psi \rightarrow \psi + h_1\lambda \Gamma_{\mu\nu}\psi
\,F^{\mu\nu} +h_2\lambda^2 \Gamma_{\mu\nu\rho\sigma}\psi
F^{\mu\nu}F^{\rho\sigma}
+ h_3\lambda^2 \psi F^2 
\]
\[+ h_4i\lambda^2 (\overline{\psi}\Gamma^{\mu\nu\rho}
\psi)\Gamma_{\mu\nu}\partial_{\rho}\psi
+h_5i\lambda^2 (\overline{\psi}\Gamma^{\mu}\partial_{\mu}\psi)
\psi
+h_6i\lambda^2 (\overline{\psi}\Gamma^{\mu}\partial^{\nu}\psi)
\Gamma_{\mu\nu}\psi\]
\EQ
+h_7i\lambda^2(\overline{\psi}\Gamma^{\mu\nu\rho}\partial_{\rho}
\psi)\Gamma_{\mu\nu}\psi+h_8i\lambda^2
(\overline{\psi}\Gamma^{\mu\nu\rho\sigma\tau}
\partial_{\tau}\psi)\Gamma_{\mu\nu\sigma\rho}\psi .
\label{fieldredefinition}
\EN
We also consider the field definition of the 
gauge field 
\EQ
A_{\mu} \rightarrow A_{\mu}+ h_9i\lambda^2(\overline{\psi}
\Gamma_{\mu\nu\rho}\psi)F^{\nu\rho}
\label{fieldredef2} .
\EN
In these definitions, the factor $i$ is inserted so that 
all the coefficients are purely real. The following formulae
for the  transformation of the coefficients under the above  
field redefinition are 
obtained by performing Fierz rearrangements appropriately. 
\EQAN
b_0 &\rightarrow& b_0-2h_1 , \\
b_1 &\rightarrow& b_1-2h_2-h_9+h_1^2 ,\\
b_2 &\rightarrow& b_2-2h_2 +b_0h_1 ,\\
b_3 &\rightarrow& b_3-\frac12 h_1^2 + h_2 ,\\
b_4 &\rightarrow& b_4 +4h_1^2-4b_0h_1 ,\\
b_5 &\rightarrow& b_5+ h_1^2+h_3 , \\
c_1 &\rightarrow& c_1 -8h_4-h_5+312h_8 ,\\
c_2 &\rightarrow& c_2 +4h_4+h_6-72h_8 , \\
c_3 &\rightarrow& c_3 -8h_4-h_6+120h_8 ,\\
c_4 &\rightarrow& c_4 -4h_4+h_6+2h_7-24h_8 ,\\
c_5 &\rightarrow& c_5 +h_7-24h_8 .
\EQNN
We note that, to the present order of the 
expansion, the field redefinition, performed for each 
chiral sector separately, does not 
affect the terms of the action
(\ref{expandedaction1}) which are the product of two chiral
sectors.  Since there are only 8 free parameters 
for the field redefinition (in each chiral sector), there
exist  three independent combinations of the coefficients
that are invariant under these transformations:
$I_1\equiv 4(b_2+2b_3)+b_0^2, \, I_2\equiv b_4-b_0^2, \, 
I_3\equiv c_2+{2\over 3}c_3-{1\over 3}c_4+{2\over 3}c_5.$ 
As long as these invariants 
are preserved, we can freely choose 8 coefficients 
out of 11 coefficients in the general form of the action. 
The nonvanishing coefficients in the original form
(\ref{expandedaction1}) of the action are 
$a_1=-1/32, a_2=1/8, b^{\pm}_0=\mp 1/2, 
b^{\pm}_4=1/2, b^{\pm}_5=1/8, c_1^{\pm}=c_2^{\pm}=1/8,
c_3^{\pm}=-1/4.$  Here the upper indices $\pm$ denote the 
chiralities of the corresponding terms. 
The values of the invariants are thus 
$I^\pm_1=1/4, \, I^\pm_2=1/4,\, I^\pm_3=-1/24.$ 
To facilitate comparison with string 
scattering amplitudes in the next section, we
perform the  field redefinition such that there remain only
the
$c_2$ terms for the quartic fermion terms 
containing a single chiral sector, by setting 
$c_1=c_3=c_4=c_5=0$ and $c_2=-1/24$.  For the
$b$-coefficients, we choose 
$b_0=b_1=0$ which leads to $b_4=1/4$. 
This choice is convenient 
in making connection to the
usual $N=1$ supersymmetry of the GSO projected action.  
Using the remaining degrees of freedom, we can further set
$b_2=1/16, b_3=b_5=0$. The coefficients of the field
redefinition are given as 
$h_1^\pm=\mp1/4, h_2=1/32, h_3=-3/16, h_4=-1/48, 
h_5=7/24, h_6=-1/12, h_7=h_8=h_9=0$.  Note that the values 
indicated without the chirality indices are common to 
both sectors.

The effective action now takes the 
form
\[
S_{{\it eff}}+{1\over \lambda^2}\int d^{10}x= 
\int d^{10}x \, 
\Big(-{1\over 4}F_{\mu\nu}^2 +{i\over 2}(\overline{\psi}_+
\Gamma^{\mu}\partial_{\mu}\psi_+ +
\overline{\psi}_-
\Gamma^{\mu}\partial_{\mu}\psi_-) \]
\[
-{\lambda^2\over 32}(F_{\mu\nu}^2)^2
+{\lambda^2\over 8}F^{\mu\nu}F_{\nu\alpha}F^{\alpha\beta}
F_{\beta\mu} \]
\[
+i{\lambda^2\over 4}(\overline{\psi}_+
\Gamma_{\mu}\partial_{\nu}\psi_+ +
\overline{\psi}_-
\Gamma_{\nu}\partial_{\mu}\psi_-)F^{\mu\alpha}F_{\alpha}^{
\,\, \nu}+
{i\lambda^2\over 16}(\overline{\psi}_+\Gamma^{\mu\nu\rho}\psi_+ +
\overline{\psi}_-\Gamma^{\mu\nu\rho}\psi_-)
F_{\mu\sigma}\partial^{\sigma}F_{\nu\rho}
\]
\[
+{\lambda^2\over 4}\overline{\psi}_+
\Gamma^{\mu}\partial_{\mu}\psi_+ 
\overline{\psi}_-
\Gamma^{\nu}\partial_{\nu}\psi_-
-{\lambda^2\over 24}(\overline{\psi}_+\Gamma_{\mu}\partial_{\nu}
\psi_+\overline{\psi}_+\Gamma^{\mu}\partial^{\nu}\psi_+
+\overline{\psi}_-\Gamma_{\mu}\partial_{\nu}
\psi_-\overline{\psi}_-\Gamma^{\mu}\partial^{\nu}\psi_-
)\]
\EQ
-{\lambda^2\over 4}\overline{\psi}_+\Gamma_{\mu}\partial_{\nu}
\psi_+\overline{\psi}_-\Gamma^{\mu}\partial^{\nu}\psi_-
+ \mbox{higher orders}
\Big) .
\label{expandedaction2}
\EN
Because of the field redefinition, the supertransformation 
law is also  changed to 
\EQ
\delta \psi\equiv \sum_{n=-1} \lambda^n\delta \psi^{(n)} , 
\quad \delta A_{\mu}\equiv \sum_{n=0}\lambda^n \delta
A_{\mu}^{(n)}
\EN
where
\EQ
\delta \psi^{(-1)}_{\pm}=\epsilon_{\pm} ,
\EN
\EQ
\delta \psi^{(0)}_\pm= \pm{1\over
4}\Gamma^{\mu\nu}\epsilon_{\pm} F_{\mu\nu} ,
\EN
\[
\delta \psi^{(1)}_\pm=
-{i\over 2}(\overline{\epsilon}\Gamma^{\mu}\psi)\partial_{\mu}
\psi_{\pm}+
{1\over16}\Gamma^{\mu\nu}\Gamma^{\rho\sigma}
\epsilon_\pm F_{\mu\nu}F_{\rho\sigma}
-{1\over 32}\Gamma^{\mu\nu\rho\sigma}\epsilon_\pm
F_{\mu\nu}F_{\rho\sigma}
+{3\over 16}\epsilon_\pm F^2
\]
\[
 \pm{i\over
4}\Gamma^{\mu\nu}\psi_\pm 
(\overline{\epsilon}\Gamma_{11}\Gamma_{[\nu}\partial_{\mu]}
\psi)
+{i\over 24}\Gamma_{\mu\nu}\partial_{\rho}\psi_\pm
(\overline{\psi}_\pm\Gamma^{\mu\nu\rho}\epsilon_\pm)\]
\[
-{7i\over 24 }\psi_\pm(\overline{\epsilon}_\pm\Gamma^{\mu}
\partial_{\mu}\psi_\pm)
-{7i\over 24}\epsilon_\pm(\overline{\psi}_\pm\Gamma^{\mu}
\partial_{\mu}\psi_\pm)
\]
\EQ
+{i\over 12}\Gamma^{\mu\nu}\psi_\pm 
(\overline{\epsilon}_\pm\Gamma_{\mu}\partial_{\nu}\psi_\pm)
+{i\over 12}\Gamma^{\mu\nu}\epsilon_\pm 
(\overline{\psi}_\pm\Gamma_{\mu}\partial_{\nu}\psi_\pm) ,
\EN
\[
 \ldots etc. 
\]
\EQ
\delta A_{\mu}^{(0)}={i\over 2}\overline{\epsilon}\Gamma_{11}
\Gamma_{\mu}\psi ,
\EN
\EQ
\delta A_{\mu}^{(1)} =-{i\over 8}\overline{\epsilon}
\Gamma_{\mu}\Gamma^{\rho\sigma}\psi
F_{\rho\sigma}-
{i\over 2}(\overline{\epsilon}\Gamma^{\nu}\psi)\partial_{\nu}
A_{\mu}-i{\lambda \over 2}(\overline{\epsilon}\Gamma^{\nu}
\partial_{\mu}\psi)
A_{\nu} ,
\EN
\[
...etc. 
\]
With this choice of the field redefinition, 
the action does not have the order $\lambda$ term.  
Correspondingly, the supertransformation 
law reduces, by dropping one of the two chiral sectors $\psi_-
=\epsilon_-=0$,
to the ordinary linear transformation in the 
$\lambda\rightarrow
0$  limit, up to the trivial fermion translation 
$\delta^{(-1)}\psi_+$.  The action, restricted 
to  a single chiral sector, coincides with the
action  given in \cite{metrah}, as it should. Equivalently, 
the supercurrents take the form 
\[
S^{\mu}_{\, \pm}={1\over \lambda}\Gamma^{\mu}\psi_{\pm}
\mp{1\over 4}\Gamma^{\alpha\beta}\Gamma^{\mu}
\psi_{\pm}F_{\alpha\beta} + O(\lambda), 
\]
where the second term restricted to a single 
chirality coincides with the ordinary
supercurrent of  the free linear theory. 
Note that, in the free theory, both terms of this 
supercurrent are conserved separately. 
As is well
known \cite{metrah}\cite{bergs}, even if we confine 
ourselves to the usual GSO projected sector, the
supertransformation law is not linear for nonzero
slope parameter ($\lambda\ne 0, \alpha'\ne 0$) even
off-shell. This is somewhat mysterious from the viewpoint of
the  usual
$N=1$ supersymmetry, because the mass spectrum after the
usual GSO projection exhibits the symmetry at each  mass
level. From our viewpoint, however, this is not surprising,
since the symmetry governing the  action is in fact the
nonlinear $N=2$ supersymmetry rather  than the $N=1$
supersymmetry.    We also note that the order $\lambda^{-1}$
terms   of the supertransformation law and the supercurrents
are just consistent  with the apparent violation of the
conservation of supercharges for matrix elements
involving a flip of  G-parity, which is, 
as elucidated in \cite{yo}, caused by the 
`wrong' boundary condition in the world-sheet
formulations. 

\section{Four-fermion scattering without the GSO projection}
\setcounter{equation}{0}

Let us now proceed to investigate the low-energy behavior 
of string scattering for the would-be Goldstone 
fermions in open string theory without the GSO projection.
Fortunately, the general expression for the open string
amplitudes with four fermion external lines  was given
longtime ago in
\cite{schwu}. Since the GSO projection \cite{gso} made the
inclusion of both chiral sectors unnecessary, the properties
of the four fermion amplitudes involving  both chiralities
in 10 dimensions have not been fully discussed in the
literature. 

We start from summarizing the old construction. 
Up to the normalization, 
the formula for $(s, t)$ amplitude with 4 massless fermion lines
is 
\EQ
A_{1,2,3,4}(s, t)=\int_0^1 dx\, x^{-\alpha's -3/2}
(1-x)^{-\alpha' t-3/2}\sum_{\ell=0}^{10} 
T^{(\ell)}(1,2)T_{(\ell)}(3,4)
f_{\ell}(x)
\label{genefermiamp}
\EN
where 
\EQ
T^{(\ell)}(1,2)=\overline{u}_1\Gamma^{(\ell)}u_2 , \, \, 
T_{(\ell)}(3,4)=\overline{u}_3\Gamma_{(\ell)}u_4, 
\EN
\EQ
f_{\ell}(x)={1\over 32}x^{5-{\ell \over 2}}
(1-\sqrt{1-x})^{\ell-5}={1\over 32}x^{{\ell\over 2}}
(1+\sqrt{1-x})^{5-\ell} .
\EN
Apart from the momentum dependent factor 
$x^{-\alpha's}(1-x)^{-\alpha't}$, the integrand 
is proportional to the vacuum expectation value of the 
product of four spin fields 
$\langle S_{\alpha}(\infty)S_{\beta}(1)
S_{\gamma}(x)S_{\delta}(0)\rangle$ multiplied by the product
of the spinor wave functions $u_{1\delta}
u_{2\gamma}u_{3\beta}u_{4\alpha}$ and by the ghost 
factor $[x(1-x)]^{-1/4}$.  The
$\Gamma^{(\ell)}$'s  (and the lower-index  counter part
$\Gamma_{(\ell)}$'s) are the normalized antisymmetric 
products of $\ell$ Gamma matrices, and
the  sum with respect to $\ell$ in (\ref{genefermiamp}) is
over the complete set  of $\Gamma^{(\ell)}$. Note that
the Lorentz indices and their contractions are 
suppressed. 
With this notation, the completeness  relation is expressed
as the identity 
\[
T^{(0)}(1,4)T_{(0)}(3,2)={1\over 32}\sum_{\ell=0}^{10}
T^{(\ell)}(1,2)T_{(\ell)}(3,4) ,
\]
which is valid for arbitrary four spinor wave functions $u_i
\, \, (i=1,2,3,4)$.  The spinor bilinears
$T^{(\ell)}(1,2)$ have symmetry  property, 
$T^{(\ell)}(1,2)=-(-1)^{{1\over 2}\ell(\ell+1)}
T^{(\ell)}(2,1)$, under the exchange of external lines. 
We also have $T^{(\ell)}(1,2)T_{(\ell)}(3,4) =\pm (-1)^{\ell}
T^{(10-\ell)}(1,2)T_{(10-\ell)}(3,4)
$ where $\pm$ depending on whether 
the lines $1$ and $3$ have the same or opposite chiralities.  
The Mandelstam variables are  defined as
$s=-(k_1+k_2)^2=-2k_1k_2, t=-2k_2k_3$  and $u=-2k_1k_3$. 

As a first exercise, let us check the case of a single
chirality where the above general formula must give the 
well known expression which is familiar from 
text books. If all the four external 
massless fermions have the same chirality, say, $+$, 
the formula is equal to 
\[
A_{1+,2+,3+,4+}(s,t)=\int_0^1 dx\, x^{-\alpha's -3/2}
(1-x)^{-\alpha' t-3/2}[
T^{(1)}(1_+,2_+)T_{(1)}(3_+,4_+)
(f_{1}(x)-f_9(x))\]
\EQ\hspace{4cm} + T^{(3)}(1_+,2_+)T_{(3)}(3_+,4_+)
(f_3(x)-f_7(x))] .
\label{singlechiralamp0}
\EN
Here and in what follows we put the indices $\pm$ 
to denote the chirality of the external lines. 
By using the identity
\[
T^{(3)}(1_+,2_+)T_{(3)}(3_+,4_+)=-2T^{(1)}(1_+,2_+)
T_{(1)}(3_+,4_+)-
4T^{(1)}(1_+,4_+)T_{(1)}(3_+,2_+), 
\]
it is easy to check that (\ref{singlechiralamp0}) reduces to
\[
\hspace{-4.8cm} A_{1+,2+,3+,4+}(s,t)={1\over
2}{\Gamma(-\alpha's)\Gamma(-\alpha't)
\over \Gamma(1-\alpha's-\alpha' t)}
\]
\EQ
\quad \times \Big(-\alpha'
tT^{(1)}(1_+,2_+)T_{(1)}(4_+,3_+) +\alpha' s T^{(1)}(1_+,4_+)
T_{(1)}(2_+,3_+)\Big) .
\label{singlechiralamp}
\EN
This is the standard form of the 4 massless fermion 
amplitude of open superstring theory 
with the ordinary GSO  projection. The kinematical
factor  in the parenthesis in eq. (\ref{singlechiralamp}),
which is  usually denoted by $-K(u_1, u_2, u_3, u_4)$ 
\cite{schre}, has 
total antisymmetry in external lines, guaranteeing the 
$s$-channel-$t$-channel duality 
$A_{1+,2+,3+,4+}(s,t)=-A_{4+,1+,2+,3+}(t,s)$. The 
minus sign shows the compatibility of the $s$-$t$ duality
with fermi antisymmetry.  In the
form  (\ref{singlechiralamp0}) which is more 
convenient for identifying the states 
propagating in each channel, the duality can be
expressed  by  the matrix identity 
\EQ
F^TM =-1 , \, \quad F=\pmatrix{-1/2 & -1/4 \cr
      -3 & 1/2}, \, \, M=\pmatrix{1/2 & 3 \cr
      1/4 & -1/2} ,
\label{singlechiraldual}
\EN
where the matrices $F, M$ are defined, respectively,  as
\EQ
\pmatrix{T^{(1)}(1_+,2_+)T_{(1)}(3_+, 4_+) \cr
       T^{(3)}(1_+,2_+)T_{(3)}(3_+, 4_+)}
=F \pmatrix{T^{(1)}(1_+,4_+)T_{(1)}(3_+,
2_+) 
\cr T^{(3)}(1_+,4_+)T_{(3)}(3_+, 2_+)}
\EN
and, with $y=1-x$, 
\EQ
\pmatrix{f_1(x)-f_9(x) \cr
        f_3(x)-f_7(x)}
= M 
\pmatrix{f_1(y)-f_9(y) \cr
f_3(y)-f_7(y) } .
\EN
The consistency of the amplitude (\ref{singlechiralamp})
with the effective  action (\ref{expandedaction2})
restricted to the single chiral sector has been checked in
\cite{metrah}. 

We now consider amplitudes with mixed chiralities. 
Let us start from the configuration with the 
ordering $(1+, 2+, 3-,
4-)$. 
\[
A_{1+,2+,3-,4-}(s,t)
=
\int_0^1 dx\, x^{-\alpha's -3/2}
(1-x)^{-\alpha' t-3/2}[
T^{(1)}(1_+,2_+)T_{(1)}(3_-,4_-)
(f_{1}(x)+f_9(x))\]
\EQ\hspace{2.5cm} + T^{(3)}(1_+,2_+)T_{(3)}(3_-,4_-)
(f_3(x)+f_7(x)) +
T^{(5)}(1_+, 2+)T_{(5)}(3_-, 4_-)f_5(x)] .
\label{twochiralamp1}
\EN
The matrix identity expressing the $s$-$t$-duality in this
case is 
\EQ
F^TM=1, \quad 
F=\pmatrix{10/16 & 6/16 & 2/16 \cr
         		120/16 & 8/16 & -8/16 \cr
  				   252/16 & -28/16 &12/16 } , \quad 
M=\pmatrix{1/16 & 27/16 & 42/16 \cr
								1/16 & 3/16 & -14/16 \cr
								1/32 & -5/32 & 10/32} ,
\label{twochiraldual1}
\EN
where the matrices $F$ and $M$ are defined  by 
\EQ
\pmatrix{T^{(1)}(1_+,2_+)T_{(1)}(3_-, 4_-) \cr
									T^{(3)}(1_+,2_+)T_{(3)}(3_-, 4_-)  \cr
        T^{(5)}(1_+,2_+)T_{(5)}(3_-, 4_-)}
=F
\pmatrix{T^{(0)}(1_+,4_-)T_{(0)}(3_-, 2_+) \cr
									T^{(2)}(1_+,4_-)T_{(2)}(3_-, 2_+)  \cr
        T^{(4)}(1_+,4_-)T_{(4)}(3_-, 2_+)} ,
\EN
\EQ
\pmatrix{f_1(x)+f_9(x)\cr
									f_3(x)+f_7(x)\cr
							f_5(x)}
=M
\pmatrix{f_0(y)-f_{10}(y)\cr
							f_2(y)-f_8(y)\cr
								f_4(y)-f_6(y)}.
\EN
Thus the amplitude (\ref{twochiralamp1}), in which 
the $s$-channel poles are the ordinary GSO projected 
states, 
is dual to $A_{1+, 4-, 3-,2+}(t,s)$ :
\EQ
A_{1+,2+,3-,4-}(s,t)=A_{1+,4-,3-,2+}(t,s) ,
\EN
\[
A_{1+,4-,3-,2+}(t,s)=
\int_0^1 dx\, x^{-\alpha't -3/2}
(1-x)^{-\alpha' s-3/2}[
T^{(0)}(1_+,4_-)T_{(0)}(3_-,2_+)
(f_{0}(x)-f_{10}(x))\]
\EQ\hspace{2cm} + T^{(2)}(1_+,4_-)T_{(2)}(3_-,2_+)
(f_2(x)-f_8(x)) +
T^{(4)}(1_+, 4_-)T_{(4)}(3_-, 2_+)(f_4(x)-f_6(x))], 
\label{twochiralamp2}
\EN
in which the $t$-channel poles are oppositely GSO
projected, as it should be. 

On the other hand, for the ordering 
$(1+, 3-, 2+,4-)$,  the  amplitude is 
\[
A_{1+,3-,2+,4-}(u,t)=
\int_0^1 dx\, x^{-\alpha'u -3/2}
(1-x)^{-\alpha' t-3/2}[
T^{(0)}(1_+,3_-)T_{(0)}(2_+,4_-)
(f_{0}(x)+f_{10}(x))\]
\EQ\hspace{2cm} + T^{(2)}(1_+,3_-)T_{(2)}(2_+,4_-)
(f_2(x)+f_8(x)) +
T^{(4)}(1_+, 3_-)T_{(4)}(2_+, 4_-)(f_4(x)+f_6(x))]. 
\label{twochiralamp3}
\EN
We can check that the duality relation is now 
\EQ
A_{1+,3-,2+,4-}(u,t)=A_{1+,4-,2+,3-}(t,u) ,
\label{mixedduality2}
\EN
\[
A_{1+,4-,2+,3-}(t,u)=
\int_0^1 dx\, x^{-\alpha't -3/2}
(1-x)^{-\alpha' u-3/2}[
T^{(0)}(1_+,4_-)T_{(0)}(2_+,3_-)
(f_{0}(x)+f_{10}(x))\]
\EQ\hspace{2cm} + T^{(2)}(1_+,4_-)T_{(2)}(2_+,3_-)
(f_2(x)+f_8(x)) +
T^{(4)}(1_+, 4_-)T_{(4)}(2_+, 3_-)(f_4(x)+f_6(x))]. 
\label{twochiralamp4}
\EN
In this case, both $u$ and $t$ channel poles are 
oppositely GSO projected. 
The matrix identity is 
\EQ
F^TM=1, \quad 
F=\pmatrix{1/16 & 1/16 & 1/16 \cr
         		45/16 & 13/16 & -3/16 \cr
  				   210/16 & -14/16 &2/16 } , \quad 
M=\pmatrix{1/16 & 45/16 & 210/16 \cr
								1/16 & 13/16 & -14/16 \cr
								1/16 & -3/16 & 2/16} ,
\label{twochiraldual2}
\EN
where  
\EQ
\pmatrix{T^{(0)}(1_+,3_-)T_{(0)}(2_+, 4_-) \cr
									T^{(2)}(1_+,3_-)T_{(2)}(2_+, 4_-)  \cr
        T^{(4)}(1_+,3_-)T_{(4)}(2_+, 4_-)}
=F
\pmatrix{T^{(0)}(1_+,4_-)T_{(0)}(2_+, 3_-) \cr
									T^{(2)}(1_+,4_-)T_{(2)}(2_+, 3_-)  \cr
        T^{(4)}(1_+,4_-)T_{(4)}(2_+, 3_-)} ,
\label{matrixidt3}
\EN
\EQ
\pmatrix{f_0(x)+f_{10}(x)\cr
									f_2(x)+f_8(x)\cr
							f_4(x)+f_6(x)}
=M
\pmatrix{f_0(y)+f_{10}(y)\cr
							f_2(y)+f_8(y)\cr
								f_4(y)+f_6(y)} .
\EN 
The duality relation (\ref{mixedduality2}) tells us a 
remarkable fact that  the duality symmetry of the
configuration
$(1+,3-,2+,4-)$  is actually contradictory to fermi antisymmetry 
under the exchange of $3-$ and $4-$ (or of $1+$ and $2+$). 
The extra minus factor under the exchange is expected 
from the extra phase factor associated to the 
OPE of fermion emission vertex operators between the
different GSO sectors.  This means that the configuration
$(1+,3-,2+,4-)$ (or its dual) does not contribute to the
total amplitude  after the summation over the inequivalent 
permutations of the external lines, which is, in general,
required to  make crossing-symmetric scattering amplitudes. 

We can provide another interpretation for the phenomenon that
the S-matrix for mixed chiralities  
does not involve the ordering $(1+,3-,2+,4-)$, 
on the basis of the property of supercharges 
in the world-sheet formulation. 
As discussed in
ref.
\cite{yo},   the necessity of inserting the Goldstone
fermions for recovering from the apparent violation of 
the conservation of 
supercharges defined on the  string world sheet 
occurs at one of the two ends of an open string. 
Namely, when we 
consider the wave function of a (+)-chirality fermion
string  regarding $(-)$-chirality fermions as external
fields,  the violation of the $N=2$ supercharges associated 
to the superparameter
$\epsilon_-$ occurs  only at one and the same end (say, 
$\sigma =\pi$) of the open string.\footnote{
See the sentence after the equation (4.9) of \cite{yo}:We have 
$dQ_{\alpha}/d\tau = -2S_{\alpha}{\rm e}^{-\phi/2}(\tau-i\pi)$, 
shifting to the Euclidean metric on the world sheet.
} Thus, the 
insertion of the Goldstone fermion, treated as external
field 
$\psi_-$, occurs only at the one end, $\sigma=\pi$, of the
(+)-chirality  fermion string. 
It is natural to assume that this property 
is satisfied for the S-matrix too. 
We can then conclude that  only the two types of cyclic
ordering,
$(1+,2+,3-,4-)$ and
$(1+,2+,4-,3-)$, are possible for 4-fermion scatterings 
with mixed chiralities.  For higher point amplitudes, 
this implies that the external lines of the
same chiralities  must always be paired as two adjacent
lines.    Of course, possible
permutations  of the external lines allowed under this
rule must be  taken into account. 
If we consider the supertransformation
associated to the  chirality
$\epsilon_+$ for the wave function of a + chirality state, 
the corresponding charge  is conserved due to the existence of
the ordinary 
$N=1$ supersymmetry without the insertion of the 
Goldstone fermions $\psi_-$. Therefore, the above
argument is not applicable. Consequently, the amplitude should
be  constructed according to the ordinary rule, leading to 
all the 3 types of cyclic ordering.

Let us next study the low-energy expansion of the 
above amplitudes. This is easily done by expressing 
the amplitudes in terms of Gamma functions. 
Taking into account 
  fermi antisymmetry, 
the end results to the order $O(\alpha')$ are
\[
\hspace{-1.8cm}
2A_{1+, 2+, 3+, 4+}(s,t)-2A_{1_+,2+,4+,3+}(s,u)
-2A_{3+,2+,4+,1+}(t,u)
\]
\EQ
= {\pi^2\over 2}\Big(\alpha'
tT^{(1)}(1_+,2_+)T_{(1)}(4_+,3_+) -\alpha' s T^{(1)}(1_+,4_+)
T_{(1)}(2_+,3_+)\Big) ,
\label{4+amp}
\EN
\[
\hspace{-5.5cm}
2A_{1+,2+,3-,4-}(s,t)-2A_{1+, 2+, 4-, 3-}(s, u)
\]
\EQ
={\pi^2\over 2}\alpha'(t-u)
T^{(1)}(1_+,2_+) T_{(1)}(3_-, 4_-) .
\label{2+2-amp}
\EN
Note that both the amplitudes (\ref{4+amp}) and 
(\ref{2+2-amp}) are normalized in the 
same way so that the massless  poles of each planar amplitude before the summation over noncyclic permutations have  residues of the same strength.  Of course, 
these massless poles corresponding to the gauge boson exchange are canceled by the fermi antisymmetrization in the 
present U(1) case.  

For comparison,
we show the result for the ordering $(1+,3-,2+,4-)$:
\[
\hspace{-5cm}
A_{1+, 3-,2+,4-}(u,t)=A_{1+,4_-, 2+,3-}(t,u)
\]
\[
\hspace{-1cm}
=\Big[-{3\alpha'\pi\over
2}s -{5\pi\over 16}(1-\alpha'sK)\Big]T^{(0)}(1_+,3_-)
T_{(0)}(2_+,4_-)
\]
\[
+\Big[-{\alpha'\pi\over 2}s
+{3\pi\over 16}(1-\alpha'sK)\Big]T^{(2)}(1_+,3_-)
T_{(2)}(2_+,4_-)\]
\EQ
-{\pi\over 16}(1-\alpha'sK)T^{(4)}(1_+,3_-)
T_{(4)}(2_+,4_-) ,
\label{zeroslopelimit3}
\EN
where $K=2(1-\ln 2)$. The difference of 
(\ref{zeroslopelimit3}) from the  previous two is that there
remain the order $(\alpha')^0$ contact terms corresponding to
the 4-fermi interaction terms
$(\overline{\psi}\Gamma^{(\ell)}\psi)^2$ ($\ell=0,2,4$) 
with no derivative. Such terms, being nonvanishing 
in the zero-momentum limit, would
contradict the nonlinear supersymmetry. However, as
emphasized above, these  terms do not contribute to the total
amplitude  after the summation over all inequivalent
(anti)permutations, due to the duality symmetry.\footnote{
It is a good  exercise to explicitly check that the
expression (\ref{zeroslopelimit3})  has the duality symmetry
which contradicts the fermi antisymmetry, using the identity
(\ref{matrixidt3}).}  

    From the viewpoint of  ordinary
field theory, this phenomena is quite miraculous. 
Remember that
there is the Yukawa interaction
$\overline{\psi}\psi
\phi$ of fermions and tachyon $\phi$, corresponding 
to poles in $t$ and $u$ channels in the amplitude 
$A_{1+,2+,3-,4-}(s,t)-A_{1+, 2+, 4-, 3-}(s, u)$ for nonzero 
$\alpha'$. The contribution of tachyon exchange  
to this amplitude is 
\[
{T^{(0)}(1_+,4_-)T_{(0)}(2_+,3_-)\over \alpha't+{1\over 2}}-
{T^{(0)}(1_+,3_-)T_{(0)}(2_+,4_-)\over \alpha'u+{1\over 2}}
\]
 giving the order $(\alpha')^0$ terms with correct fermi 
antisymmetry. However, the exact amplitude 
again does not have the order $(\alpha')^0$ contribution. 
Clearly, in both this and the case (\ref{zeroslopelimit3}), 
the infinite  tower of massive modes of the oppositely
GSO-projected  states guaranteeing  the validity of the
duality relation (\ref{twochiraldual2}) is responsible for the
cancellation of the  order $(\alpha')^0$ terms in the
low-energy limit.  In \cite{yo}, a conceptually 
similar phenomenon related to 
the duality between open and closed string channels was 
pointed out for the case of the supersymmetric  mass
formula,  expressed as 
$\Tr\, [(-1)^F M^{2n}]=384\, \delta_{n\,
,4}/(2\pi\alpha')^4$.  These  properties should be kept in
mind as an indication that  it is essential to take into
account all the  massive modes, when we discuss the hidden
supersymmetry for  nonBPS D-branes.

Finally,  comparing the amplitude(s) (\ref{2+2-amp}) (and 
(\ref{4+amp}) which has already been checked in \cite{metrah}) with the
effective action (\ref{expandedaction2}), we find that  the
open string amplitudes with the both GSO 
 sectors indeed satisfy  the
low-energy theorem of the nonlinear 
$N=2$ supersymmetry, by relating the necessary 
overall normalization constant $g^2$ for 
(\ref{2+2-amp}) and (\ref{4+amp}) to the Yang-Mills 
coupling $\lambda$ as 
$g^2\alpha'\pi^2/2=\lambda^2/4\,
(=g_s(2\pi)^9\alpha'^5/4)$.  Note that only the last term
in the effective action (\ref{expandedaction2}) 
contributes to the 4-point on-shell S-matrix elements with
mixed chiralities.  The validity of the effective action
for the other terms involving gauge field is already
known, since those terms are essentially the same as in
the  case with the usual GSO projection.

\section{Concluding remarks}

We studied the low-energy effective action 
for open strings without the GSO projection, by assuming the
existence  of the nonlinearly realized $N=2$ supersymmetry
and  showed that the low-energy behavior of the 
4-point amplitudes of would-be Goldstone 
fermions 
is perfectly consistent with the hidden $N=2$ supersymmetry. 
This provides a further concrete support to earlier 
 discussions in \cite{sen} and  
in \cite{yo}.  These previous works
 have analyzed the aspects of the hidden $N=2$  
supersymmetry from mutually complementary standpoints.  

There remain many further questions 
to which we hope to return in future works. For instances:
(1) In the present paper, we have only treated the 
$U(1)$ case of type IIA theory. The type IIA 
effective action can trivially be changed to type IIB
case \cite{aps}\cite{sen} for a single 
Dp-brane.  On the other hand,  the
extensions of the effective action to the
unstable D9-$\overline{{\rm D9}}$ system of type IIB and 
 to general non-Abelian cases \cite{tseytlin}  are not so
straightforward.  On the side of the open-string
computation, we can  easily extend our results to
non-Abelian cases by taking into account the  Chan-Paton
factors appropriately.  (2) Of course, 
generalizations to higher orders in
$\lambda$-expansion  and to higher derivatives are also 
 important. It is very desirable to formulate the 
higher derivative corrections in a systematic 
geometric manner. (3) We
have considered only the  open string sector. 
This is justified in the weak string-coupling region 
($g_s \ll 1$), since 
the gravitational length $\ell_P \sim
g_s^{1/4}\sqrt{2\pi\alpha'}$ charactering 
gravity in string theory is much smaller
than the characteristic  length $\ell_{{\it susy}}\sim
\lambda^{1/5}\sim  g_s^{1/10}\sqrt{2\pi\alpha'}$ associated
with the supersymmetry breaking in the present effective
open-string theory.  Nonetheless, 
 it is desirable to extend the present work and the
discussions in the previous works \cite{sen}\cite{yo}
including  all the supergravity background fields.  If the
supergravity fields are treated dynamically as closed 
strings,  the Goldstone fermions are expected to be absorbed 
by gravitino through the super Higgs effect in the wrong 
tachyonic vacuum. 
In addition to this, there is a nonvanishing cosmological 
constant corresponding to the tension of D9-brane.  
How such effects can affect the 
dynamics of unstable systems is an interesting question. 
(4) It is important to discuss the role of the 
$N=2$ supersymmetry in the dynamics of tachyon 
condensation. 
Within the validity of the lowest nontrivial 
approximation we are using, the
supertransformation law (\ref{strpsi}) implies  that the
restoration of the full
$N=2$ supersymmetry  requires $\langle
i\overline{\psi}_{\pm \beta}\partial_{\mu}
\psi_{\pm\alpha}\rangle
=-(\Gamma_{\mu}(1\pm \Gamma_{11})/2) _{\alpha\beta}/5\lambda^2$
before the  field redefinition. A
meaningful  question we may ask here is how this vacuum
expectation  value is related to tachyon condensation. 
One possible approach might be to directly construct the 
effective  action for fermion bilinears starting from the 
effective  action (\ref{effaction}) without introducing 
tachyon fields explicitly. 
As we have emphasized repeatedly, the massive
excitations of open strings other than the tachyon play
essential roles for the hidden $N=2$ supersymmetry. We note
that a simple addition of tachyon field to the effective
action  is a rather dubious procedure, in view of 
the nature of our effective action in which all 
massive modes are integrated out at equal footing.  
Unfortunately, the standard methods for treating string
field theory,  which is the only framework in our
hand at present to treat all massive 
modes democratically, are not particularly convenient for
exhibiting supersymmetry  and other exact properties of
stringy nature such as 
$s$-$t$, open-closed string dualities and modular
invariance.  
It is desirable to develop more powerful
formalisms   for investigating the nonperturbative structure
of  string theory and underlying symmetries.

\vspace{1cm}
\noindent
Acknowledgements

One (T.Y.) of the present authors would like to 
thank his colleagues, especially, Y. Kazama,  and  participants
of SI2000 workshop (August, 2000) at Mt. Fuji for 
conversations and comments related to this work.  The present
work is supported in part by Grant-in-Aid for Scientific 
Research (No. 12440060)  from the Ministry of  Education,
Science and Culture. 
 

\end{document}